\documentclass[11pt,a4paper]{article}
\usepackage[margin=2.5cm]{geometry}
\usepackage[utf8]{inputenc}
\usepackage{amsmath,amssymb}
\usepackage{graphicx}
\usepackage[aboveskip=1pt,labelfont=bf,labelsep=period]{caption}
\usepackage{url}

\graphicspath{{./figs/}}

\date{}

\newcommand{\yk}{\mathbf{y}^{(k)}}
\newcommand{\wk}{w^{(k)}}
\newcommand{\wm}{w^{(m)}}
\newcommand{\tw}{\textwidth}

\begin{document}

\begin{flushleft}
{\Large
\textbf\newline{Prediction of employment and unemployment rates from Twitter daily rhythms in the US}
}
\newline
\\
Eszter Bokányi, bokanyi@complex.elte.hu\textsuperscript{*},\\
Zoltán Lábszki, labszkizoltan@gmail.com,\\
Gábor Vattay, vattay@complex.elte.hu\\
\bigskip
Department of Physics of Complex Systems\\
Pázmány Péter sétány 1/A, Eötvös Loránd University, Budapest H-1117, Hungary
\bigskip

* corresponding author
\end{flushleft}

\clearpage

\section*{Keywords}

unemployment prediction, Twitter, social media, activity patterns

\section*{Abstract}

By modeling macro-economical indicators using digital traces of human activities on mobile or social networks, we can provide important insights to processes previously assessed via paper-based surveys or polls only. We collected aggregated workday activity timelines of US counties from the normalized number of messages sent in each hour on the online social network Twitter. In this paper, we show how county employment and unemployment statistics are encoded in the daily rhythm of people by decomposing the activity timelines into a linear combination of two dominant patterns. The mixing ratio of these patterns defines a measure for each county, that correlates significantly with employment ($0.46\pm0.02$) and unemployment rates ($-0.34\pm0.02$). Thus, the two dominant activity patterns can be linked to rhythms signaling presence or lack of regular working hours of individuals. The analysis could provide policy makers a better insight into the processes governing employment, where problems could not only be identified based on the number of officially registered unemployed, but also on the basis of the digital footprints people leave on different platforms.

\clearpage

\section*{Introduction}

Until recently, it has been a time-consuming, costly and arduous work to collect and analyze data about individual humans at a large scale. With the advent of the digital era, there is a growing amount of data accessible online that enables the analysis and modeling of human behavior. However, our understanding of these digital data sources and the methods that connect the data to real-world outcomes is still limited.

Several aspects on the possible usage of mobile phone records and social media status updates in the estimation of official data, such as census, demographic or land use records have been discussed in recent papers. A promising approach is the analysis of the diurnal rhythm of humans. Due to the 24 hour periodicity of the Earth's rotation, we are biologically bound to show daily periodic behavior both at the individual and at the aggregate level. This periodic cycle is governed mainly by internal biochemical processes \cite{Aschoff1976,Cagnacci1992,Refinetti1992,Cajochen2003}, but the impact of external factors and the environment also leaves its imprint on these daily patterns \cite{Taillard1999,Aledavood2015}.

As S\"aramaki and Moro point out in their paper \cite{Moro2015}, an interesting application is to consider the geospatial aspects of the aggregate level of daily rhythms, as it can provide insight into several different phenomena ranging from the actual land use patterns in a city \cite{Reades2007,Reades2009,Calabrese2010,Soto2011,Toole2012,Pei2014,Louail2015,Grauwin2015,Kondor2015,Lenormand2015,Cici2015} and on a campus \cite{Calabrese2010}, to the tracking of anomalous events \cite{Candia2008,Cici2015}, or the estimation of population size \cite{Douglass2014}, mobility patterns \cite{Schneider2013}, poverty \cite{Smith-Clarke2014} or crime rates \cite{Bogomolov2014} in a certain area.

Because these aggregate patterns always consist of the superposition of the daily rhythms of individuals, it is worth investigating how the main features of the aggregate level form from superposition. 
If we can cluster individuals into more or less homogeneously behaving groups based on their daily patterns \cite{Jiang2012}, then the aggregate pattern can be understood as the combination of the group patterns, and the group that has more individuals dominates the aggregate daily rhythm. The groups of individuals can form along many demographic and/or socioeconomic factors, of which being employed and going to and from work at regular hours is the most determining one with respect to the daily activity patterns. Thus, decomposing the groups from the aggregate patterns in different geographical regions may give insight into the estimation of employment statistics in that region.

Nowcasting or estimating unemployment rates using the digital traces of search engines has already been in the focus of several papers \cite{Ettredge2005,Choi2012,Pavlicek2015}. It has already been shown, that daily activity patterns of individuals can be linked to the regularity of their working hours \cite{Eagle2006}. Because the loss of a job has severe psychological consequences \cite{Proserpio2016}, the effects of a mass layoff can be detected in the unemployment rates and provide a possibility of forecasting macro-economical effects based on observation of several individuals \cite{Toole2015}. In \cite{Llorente2015}, there is a strong evidence that aggregated daily activities of certain time intervals of geographical regions can be indicative of unemployment rates.

In this paper we obtain 63 million geolocated messages from the publicly available stream of the social network Twitter from the area of the United States sent between January and October 2014. We aggregate Monday to Friday relative tweeting activity for each hour in each US county to form an average workday activity pattern. We then assume that these activity patterns form a roughly linear subspace of the 24-hour ``timespace''. By finding this linear subspace, that is, by finding the line on which the county patterns lie, we are able to give a measure that is linked to the ratio of two groups of people tweeting in a county. We then show that this measure correlates significantly with county employment and unemployment rates, and that the average patterns corresponding to the two groups can be linked to lifestyles connected to regular working hours or the lack of them. We thus give a possible framework for decomposing the digital activity patterns of geographical regions and linking the decomposition to employment and unemployment rates.

\section*{Methods}

\subsection*{Twitter dataset}
We use the data stream freely provided by Twitter through their
Application Program Interface, which amounts to approximately 1\% of all sent
messages. In this study, we focus on the part of the data stream with 
geolocation information. These geolocated tweets originate from users who chose to allow 
their mobile phones to post the GPS coordinates along with a Twitter message. The 
total geolocated content was found to only comprise of a small percentage of all tweets;
therefore with data collection focusing only on these, a large fraction of all 
geo-tagged tweets can be gained \cite{Morstatter2013}.
Our dataset includes a total of 63 million tweets from the contiguous United
States collected between January 2014 and October 2014. These are all geotagged --
that is, they have GPS coordinates associated with them. We construct a
geographically indexed database of these tweets, permitting the efficient 
analysis of
regional features \cite{Dobos2013}. Using the Hierarchical Triangular Mesh
scheme for practical geographic indexing \cite{Szalay2007,Kondor2014},
we assigned a US county to each tweet. County borders are obtained from the
GAdm database \cite{gadm}.

\subsection*{Demographic datasets}

\label{sec:dataset}

For the population-weighed linear model of the next section, we obtain 
county-level population statistics from the US 2010 Census \cite{us_census}. We download the 
unemployment and labor force data for the time window of the Twitter dataset 
from the Local Area Unemployment Statistics page of the Bureau of Labor 
Statistics \cite{us_bls}. We take an average of the months ranging from January 2014 to October 2014 for each county. 

Though unemployment levels are defined as the number of unemployed per total labor force in a county, we define the share of employed as the number of employed divided by the whole population of a county. This measure fits the model for the daily rhythm better as discussed in the Results section.

\subsection*{Daily activity patterns}

We define a daily activity pattern with hourly resolution for each county, which 
are enumerated by $k=1\dots M$. We take all tweets originating from a given county 
from the period between January 2014 and October 2014. Then we aggregate the 
number of tweets ($n_i$) in each hour (the hour range goes from $i=0,1,\dots,23$) 
on workdays, that is from Monday to Friday, after correcting for timezone and daylight saving time in each county. Because of the differing population 
and Twitter penetration rates (share of people using Twitter) in each county, we 
normalize the number of tweets by the total number of tweets counted. Thus, each 
county ($k$) is represented by a 24-dimensional vector ($\yk$), where the 
elements of $\yk$ are:
\[y_i^{(k)}=\frac{n_i}{\sum_{i=0}^{23} n_i},\]
and obviously,
\[\sum_{i=0}^{23}y_i^{(k)}=1\quad \forall k=1\dots M.\]

To improve the quality of our dataset, we consider only those counties in which 
the overall tweet count during the ten month exceeded the threshold of 1800. Thus, we are left with 1884 counties for our analysis.

\subsection*{Linear model}

We assume that the tweeting pattern of a county can be represented by the linear combination of only two universal patterns  ($\mathbf{A}$ and $\mathbf{B}$) that are mixed for each county $k$ with a proportion of $\alpha^{(k)}$, and $1-\alpha^{(k)}$, respectively. Thus, we identify the two universal patterns that compose the pattern of a county as corresponding to two differently behaving population groups, whose aggregate tweeting patterns form $\mathbf{A}$ and $\mathbf{B}$. We have no further restriction on these $\alpha^{(k)}$ values, they can be any arbitrary real numbers.

Then the predicted activity $x_i^{(k)}$ of a county $k$ in hour $i$ would be the following linear combination:
\begin{equation}
x_i^{(k)}=\alpha^{(k)}\cdot A_i+(1-\alpha^{(k)})\cdot B_i=\alpha^{(k)}(A_i-B_i)+B_i.
\label{eq:linear}
\end{equation}

Let us denote the weight of each county by $\wk$, which is proportional to its population $p^{(k)}$, such that $\wk=p^{(k)}/\sum_{k=1}^M p^{(k)}$. We then define the squared error of our model as
\[E=\sum_{i,k}\wk\left(y_i^{(k)}-\underbrace{\left(\alpha^{(k)}(A_i-B_i)+B_i\right)}_{x_i^{(k)}}\right)^2.\]

We would like to minimize this error with subject to the two conditions $	\sum_i A_i=1, \sum_i B_i=1$. It can be shown (see SI), that the minimum occurs if $\mathbf{A-B}$ is parallel to the eigenvector $\mathbf{m}$ corresponding to the biggest eigenvalue of the weighed covariance matrix $\mathbf{C}$, and that $\mathbf{B}$ can be chosen as the average of $\yk$s. Here, an element of the covariance matrix $\mathbf{C}$ is
\begin{equation}
	\label{eq:cov}
	C_{ij}=\left\langle y_iy_j\right\rangle-\left\langle y_i\right\rangle \left\langle y_j\right\rangle,
\end{equation}
where
\begin{equation}
	\left\langle y_j\right\rangle=\sum_k \wk y_j^{(k)}.
\end{equation}

In both cases, we now consider a linear representation of the data with a coordinate system where the mean $\left\langle \mathbf{y}\right\rangle$ sets the origin and $\mathbf{m}$ is the direction of the line. We calculate $\alpha^{(k)}$ values for each county by projecting $y^{(k)}$ onto this line (see SI). A positive $\alpha^{(k)}$ means a county, where the majority of people are active on Twitter in correspondence with the daily rhythm dictated by $\mathbf{m}$, accordingly, negative $\alpha^{(k)}$ is in connection with an opposite pattern. 

Because the linear equation system derived from the minimization of the squared error is linearly dependent, the scale on our line is not set (see SI), as $\mathbf{A}-\mathbf{B}$ is only determined up to an arbitrary scaling factor. Thus, the $\alpha^{(k)}$ values are also determined only up to a scaling factor. Let us now choose $\mathbf{A}$ and $\mathbf{B}$ to be two standard deviations of $\alpha^{(k)}$-s away from the origin $\left\langle \mathbf{y}\right\rangle$ in the two directions of our new linear coordinate system:
\begin{eqnarray}
	\sigma(\alpha)=\sqrt{\frac{\sum_{k=1}^M\left(\alpha^{(k)}\right)^2}{M}}, \nonumber\\
	\mathbf{A}=\left\langle \mathbf{y}\right\rangle + 2\cdot \mathbf{m}\cdot \sigma(\alpha), \label{eq:a}\\
	\mathbf{B}=\left\langle \mathbf{y}\right\rangle - 2\cdot \mathbf{m}\cdot \sigma(\alpha). \label{eq:b}
\end{eqnarray}

$\mathbf{A}$ and $\mathbf{B}$ are both normalized to 1, where in the 2-dimensional case their components represent the selected two hours, while in the 24 dimensional case they represent the 24 hours of the day.

\label{sec:methods}

\section*{Results and discussion}

In this section, we present the description and the discussion of the main results of this paper. First, we investigate the correlation between the activities of individual hours and employment and unemployment rates, and choose two dimensions with which employment and unemployment levels have maximum or minimum correlations. We then evaluate to what extent the linear model is a valid description of our data for these most separating dimensions (2) and then for all possible dimensions (24) of our dataset. Second, we discuss how the linear models in 2 and 24 dimensions separate the two population groups with the two distinct activity patterns, and give a possible interpretation of these patterns. Third, we connect the two groups with real-world indicators like share of employed in a county, and discuss the plausibility of the correspondence of the daily patterns of the two separate groups to employment status.

\begin{figure}[h!]
	\begin{center}
		\includegraphics[width=0.7\tw]{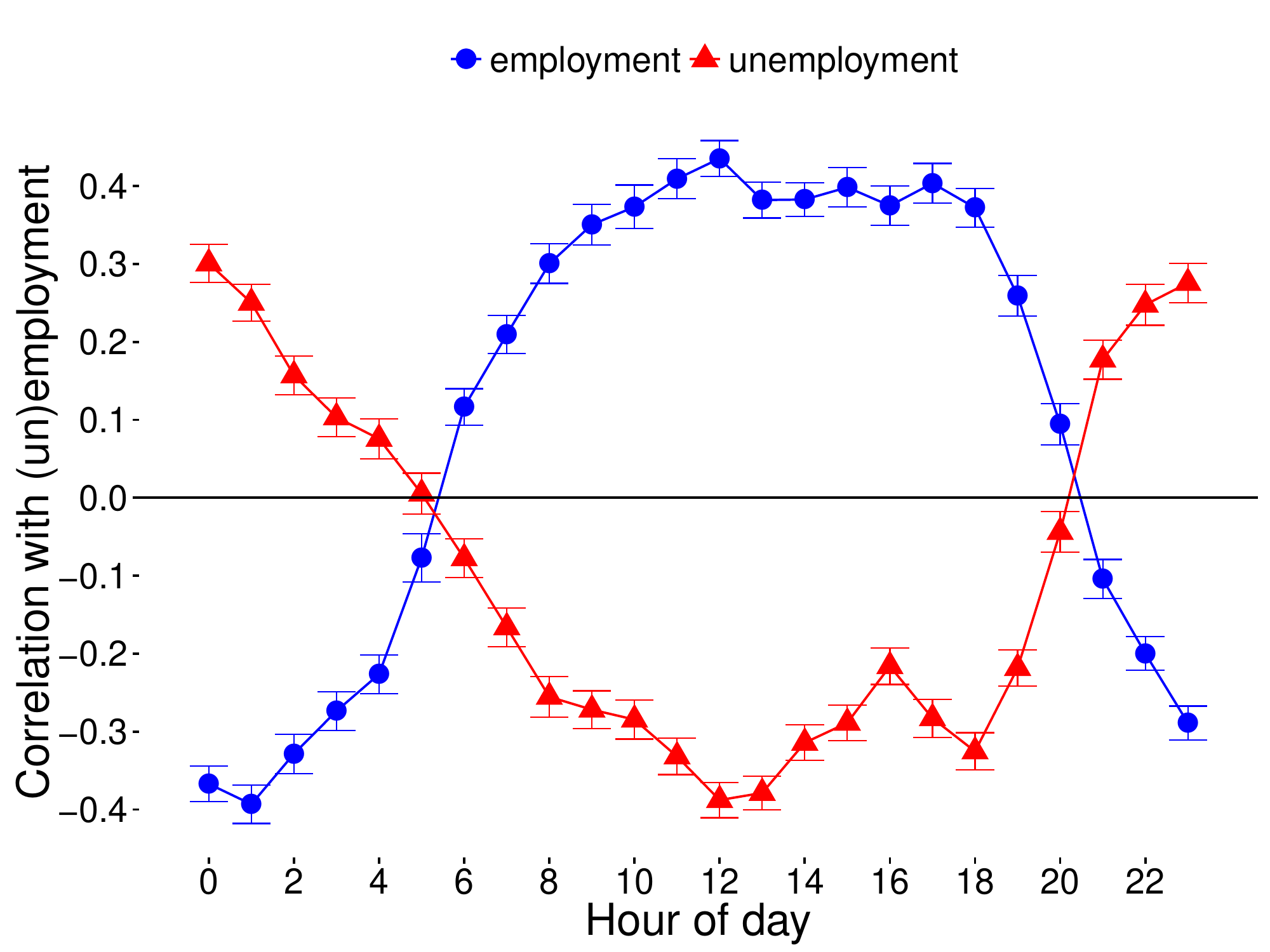}
	\end{center}
	\caption{{\bf Population-weighted Pearson correlation of employment and unemployment levels with hourly activities.} Errorbars are calculated using bootstrapping $n=1000$ times. The hours between 6am and 8pm correlate significantly positively with employment and negatively with unemployment. This relationship turns out to be exactly the opposite during the night. Regarding employment, the most distinguishing hours are 0am (most negative correlation) and 1pm (most positive correlation).}
	\label{fig:hourcorr}
\end{figure}

We first evaluate population-weighted Pearson correlations for each hour $i$ between $y_i^{(k)}$ activities for the 1884 counties (from which we have an adequate number of messages) and employment and unemployment levels. We calculate the errors of these correlations by bootstrapping our sample for $n=1000$ times, the results with errorbars are shown in Fig~\ref{fig:hourcorr}. While unemployment levels are defined in the traditional way of the Bureau of Labor Statistics, we define the share of employed slightly differently, normalizing the number of employed by the entire population of a county. This definition matches the notion of population share of ``active'' people regarding regular working hours better.

The hours between 6am and 8pm show a significantly positive correlation with employment, and a negative one with unemployment, while during the night, between 9pm and 5am, the correlation is reversed. 
With respect to employment, the correlation peaks at 12am with $0.43\pm0.02$ and reaches its lowest value at 1am with $-0.39\pm0.03$. The location of the maximum and minimum of correlation with unemployment are shifted slightly to 0am and 12am, though exactly with opposite signs ($0.30\pm0.02$ for 0am and $-0.38\pm0.02$ for 12am). The signs of the correlations and the hours of their extreme values indicate that increased daytime activity is associated with higher employment levels, and higher than average nighttime activity corresponds to higher unemployment.

\begin{figure}
	\begin{center}
		\includegraphics[width=0.7\tw]{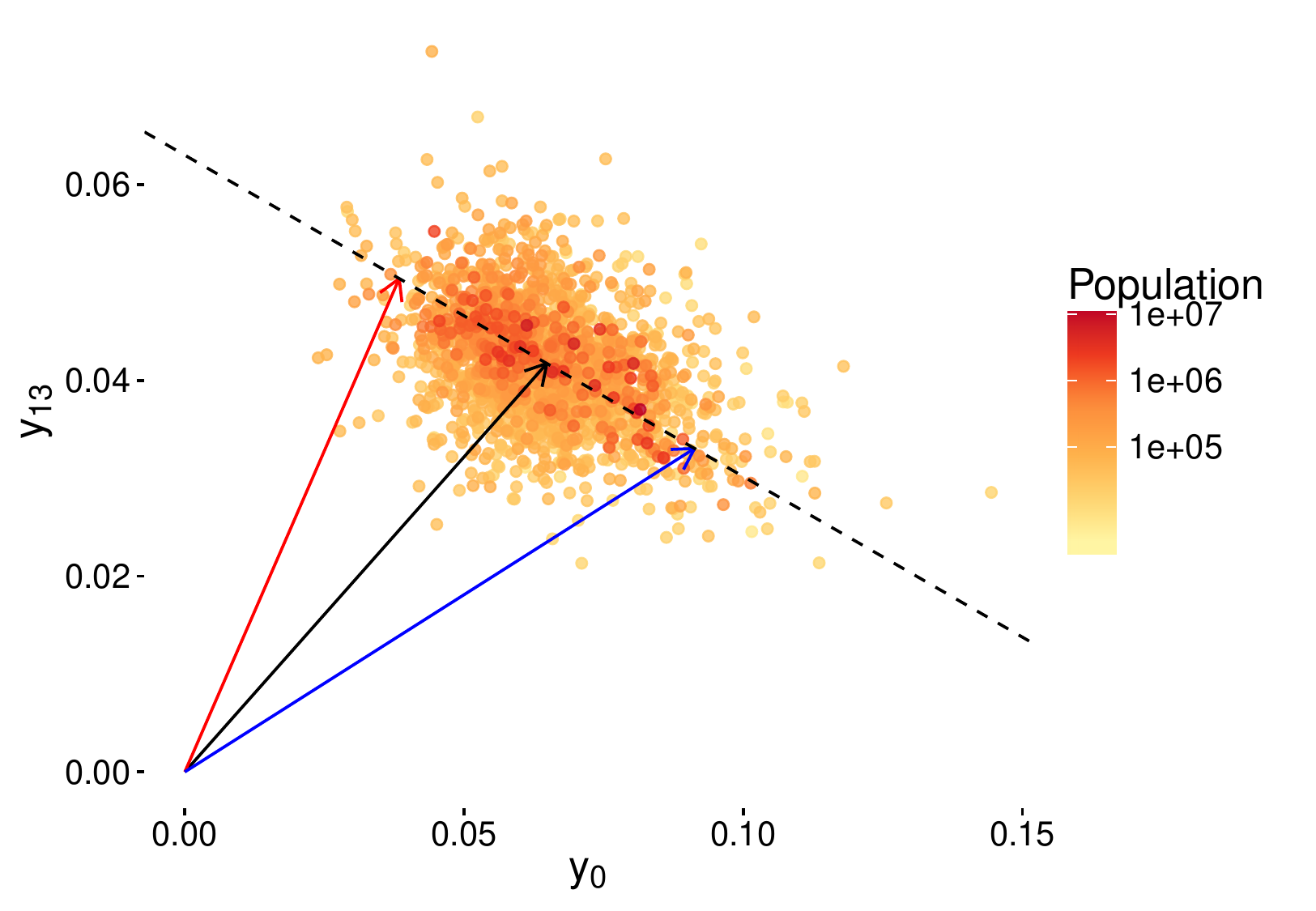}
	\end{center}
	\caption{\b{Activity of counties in the space of 0am and 1pm.} Each dot represents a county, and the horizontal axis measures the relative tweeting activity between 0am and 1am in that county, while the vertical axis represents the relative tweeting activity between 1pm and 2pm in that county. As these two measures are correlated, a linear transformation cound combine them into a single coordinate. The new coordinate axis is represented by the dashed line. The black arrow points to the average of the measures along the original axes. The blue and the red arrows are possible choices for $\mathbf{A}$ and $\mathbf{B}$ vectors, see the Linear model part in the Methods section.}
	\label{fig:two}
 
\end{figure}

To check the linearity of the model described in the Methods section, we first choose the coordinate system of the hours having the extreme correlation values with employment levels. Fig~\ref{fig:two} shows the 0am and 1pm activities of the filtered counties with the dashed line corresponding to the direction of the first eigenvector of the covariance matrix, now calculated only from these two dimensions. If we normalize the eigenvalues by their sum, we see that the first eigenvalue of the covariance matrix carries 0.99 share from all the variance in the data, thus, linearity in this two-dimensional subspace of the whole 24-hour activity space is a good assumption.

\begin{figure}
\caption{{\bf The result of the principal component analysis of the population-weighted covariance matrix.} \textbf{a} Proportion of explained variance for the principal components of the covariance matrix. Only the first four components carry a share of variance significantly greater than zero. \textbf{b} Principal component corresponding to the largest principal value. The amplitude of vector components is plotted, each vector coordinate corresponds to an hour ranging from $0$ to $23$. The vector components are positive from 5am to 8pm, and negative otherwise.}
\centering
\includegraphics[width=0.9\tw]{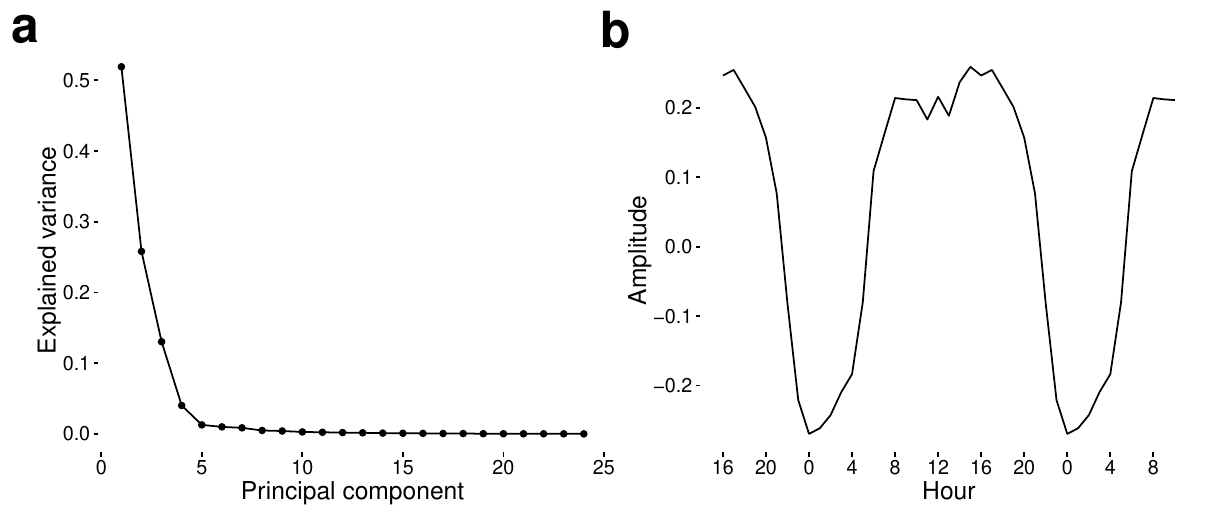}
\label{fig:prcomps}
\end{figure}

We continue by assessing the validity of the linear model in all 24 dimensions presented in Eq~\ref{eq:linear}. In Fig~\ref{fig:prcomps}a we plot eigenvalues of the covariance matrix $\mathbf{C}$ again normalized by the sum of all eigenvalues. Only the first four eigenvalues correspond to a variance significantly greater than 0, and the first principal component stands out with a proportion of 0.52, whereas the other three significant components carry 0.25, 0.13 and 0.04 share of the variance. Thus, our dataset is mostly linear even in the 24-dimensional space, and the representation with Eq~\ref{eq:linear} remains plausible.

In the 2-dimensional case, the dashed line of Fig~\ref{fig:two} marks the direction of the first principal vector. The difference between the two vectors $\mathbf{A}$ (red) and $\mathbf{B}$ (blue) representing the two universal patterns (see Methods on  p.~\pageref{sec:methods}) is parallel to this component, let us denote it by $\mathbf{m}$. It can be seen in Fig~\ref{fig:two} that the $\mathbf{A}$ pattern is marked by an increased activity at 1pm, and a decreased activity at 0am, while pattern $\mathbf{B}$ is characterized by exactly the inverse relationship.

The principal component corresponding to the largest principal value in the 24-dimensional case can be seen in Fig~\ref{fig:prcomps}. As the coordinates represent the hours, it can be seen from Fig~\ref{fig:prcomps} that $\mathbf{m}$ is positive from 5am until 8pm, and negative otherwise. Thus, the positive elements of $\mathbf{m}$ select mainly those hours during which people are awake, and the negative elements correspond to the sleeping hours.

\begin{figure}
\caption{{\bf Activity patterns corresponding to the two population groups.} The red line, $\mathbf{A}$ corresponds to the daily activity pattern of a population with regular working hours. The blue line, $\mathbf{B}$ corresponds to the other group who stay up until later in the evening and wake up later as well. The dashed line marks the average activity of all counties.}
\includegraphics[width=\tw]{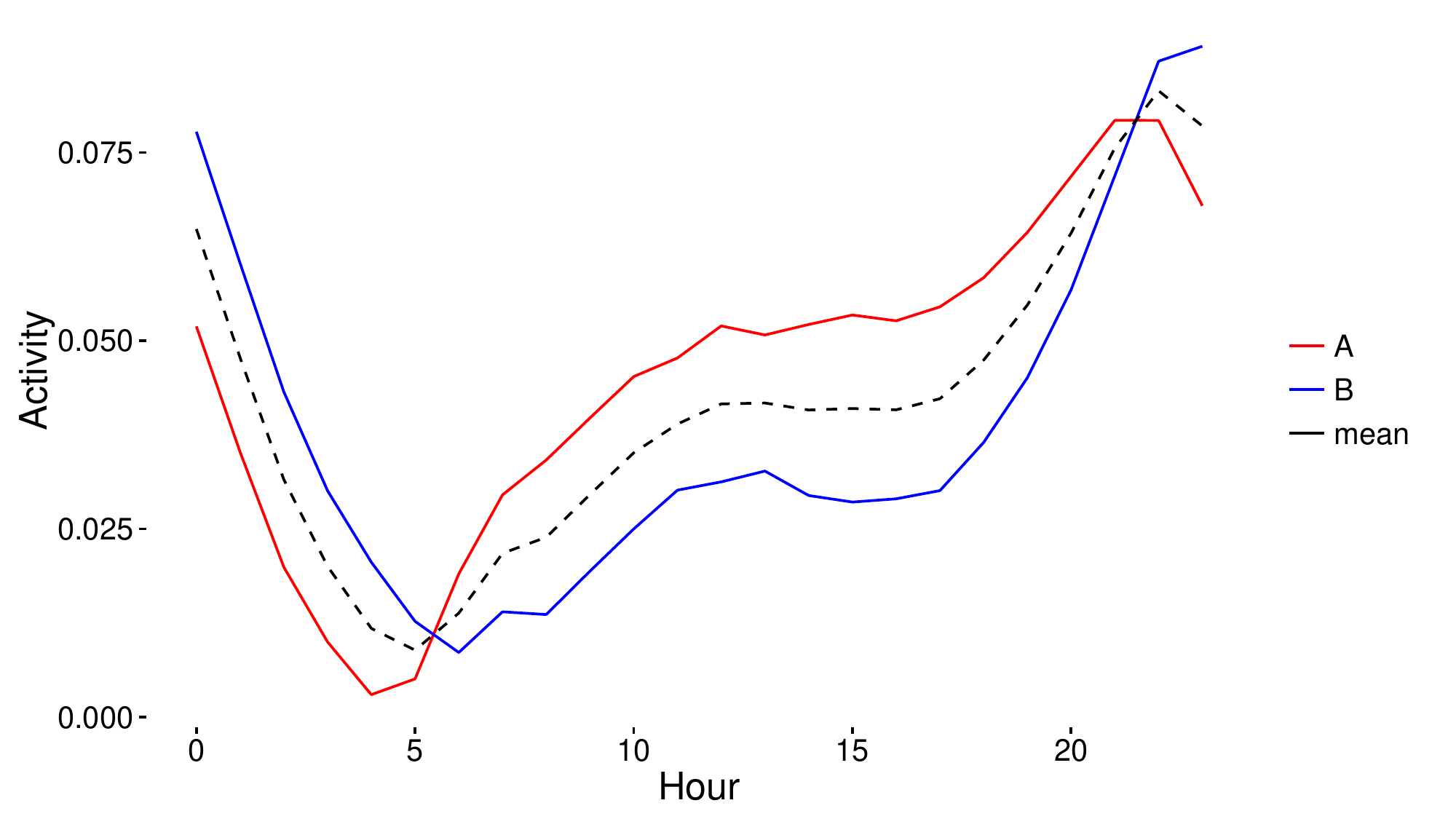}
\label{fig:ai}
\end{figure}

We then plot the elements of the 24-dimensional $\mathbf{A}$ and $\mathbf{B}$ from Eq~\ref{eq:a}-\ref{eq:b} in Fig~\ref{fig:ai}. By interpreting these patterns as the different average tweeting patterns of two population groups, each $\alpha^{(k)}$ is proportional to the share of people in a county in one population group. Our hypothesis is that the group more active during the daytime corresponds to people who regularly go to work, school etc. on weekdays, thus their daytime is regulated by the earlier wake-up and bedtime indicated in pattern $\mathbf{A}$. On the other hand, pattern $\mathbf{B}$ could correspond to a group where this regulation factor does not exist due to retirement, unemployment or any other reason, which would allow these people to be more active during nighttime and wake up later.

\begin{figure}
	\begin{center}
		\includegraphics[width=\tw]{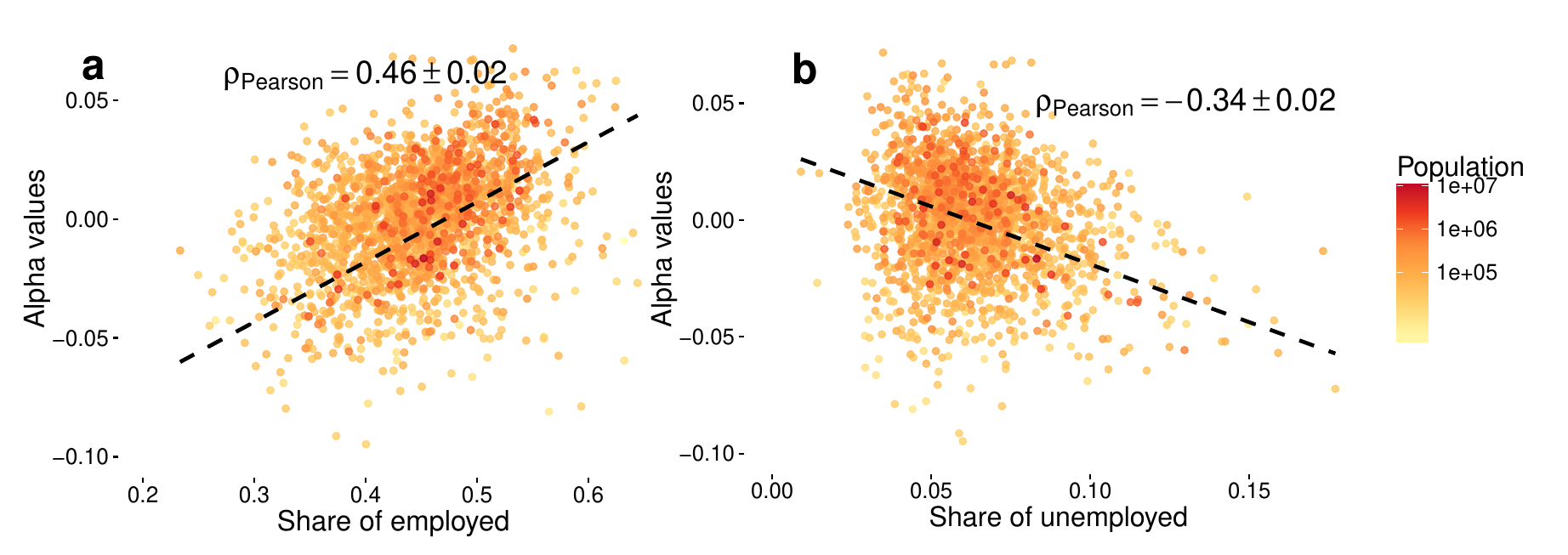}
		\caption{{\bf Scatterplots of 24-dimensional projected $\alpha^{(k)}$ values with employment and unemployment}}
		\label{fig:scatter}
	\end{center}
\end{figure}

\begin{figure}
	\begin{center}
		\includegraphics[width=\tw]{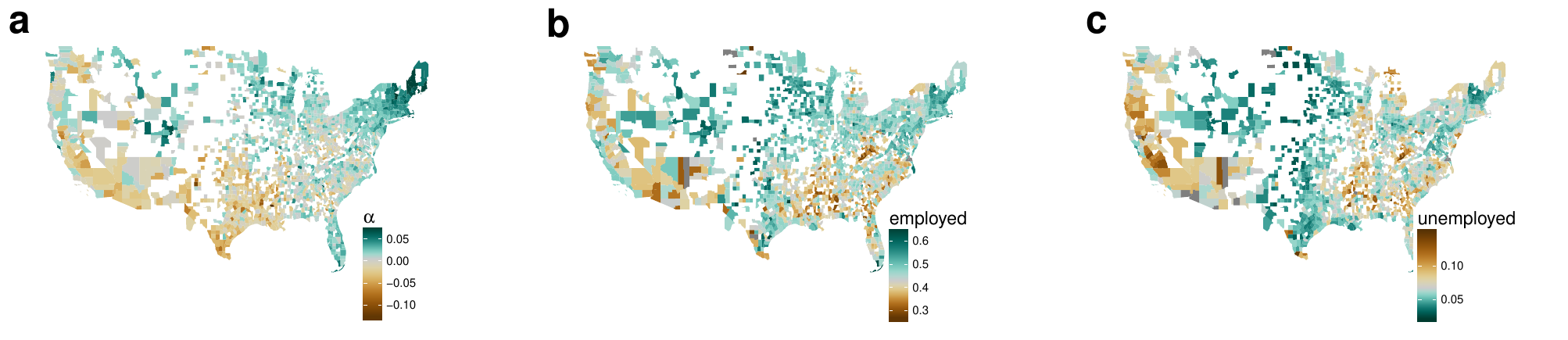}
		\caption{{\bf Map of $\alpha^{(k)}$, employment and unemployment levels.} Regional similarities are visualized by plotting \textbf{a} $\alpha^{(k)}$ measures, employment \textbf{b} and unemployment \textbf{c} on a US county map. Blank counties did not exceed the 1800 tweet threshold described in the Methods section.}
		\label{fig:maps}
	\end{center}
\end{figure}

To confirm our hypothesis, we correlate $\alpha^{(k)}$ values with labor force and unemployment estimates from the Local Area Unemployment Statistics (see Methods on p.~\pageref{sec:dataset}) of the investigated counties. In the 2-dimensional case, these combined values of $\alpha^{(k)}$ do not correlate with employment ($0.38\pm0.03$) or unemployment ($-0.32\pm0.02$) better than previous activity measures from single dimensions from Fig~\ref{fig:hourcorr}. However, by using all dimensions, we find correlations of $0.46\pm 0.02$ and $-0.34\pm 0.02$ for employment (see scatterplot in Fig~\ref{fig:scatter}) and unemployment, respectively. For the employment this is an improvement to that of the single dimensional correlations, while it is not for the unemployment. A possible interpretation is that a stricter daily rhythm is imposed upon those who are employed, as such, the characteristics of their activity curves mean a stronger overall pattern than that of the unemployed. Nevertheless, the result shows that high a $\alpha^{(k)}$ is significantly bound to higher employment, and lower unemployment rates, and that the overall shape of the activity timeline can give us more information than just using one feature of a whole day. The similarity of the regional distribution of $\alpha^{(k)}$, unemployment and employment rates are visualized on the three maps of Fig~\ref{fig:maps}.

Our results are in line with previous research carried out for Spain in \cite{Llorente2015}, where share of Twitter activity during a window of the morning hours (8-11am), afternoon hours (3-5pm) and of the night hours (0-3am) correlated significantly with unemployment rates among 25 to 44-year old inhabitants of Spanish administrative areas. High morning and low night activity indicated lower unemployment rates, which is in correspondence with our correlations. Although in Spain high afternoon activity correlated positively with unemployment levels, we cannot observe this phenomenon in the US. Due to the bias in the age of Twitter users towards younger age groups \cite{Duggan2015}, our calculated county activity patterns are not representative of the whole population. We believe that our model could be improved by incorporating labor force data detailed by different age groups.

That correlation with unemployment is significantly lower than correlation with labor force share of the population can be related to the fact that the share of employed should overlap more with the population exhibiting the ``working'' pattern $\mathbf{A}$, whereas officially registered unemployed people are not distinguishable in this context from those who are on a maternal leave or are retired etc. We also believe that there are other inherent reasons for example the more flexible working hours in the creative industry that limit the power of such a simple model explaining the employment patterns of a geographical area.

\section*{Conclusions}

In this paper we analyzed an extensive collection of geolocated tweets originating from the United States between January 2014 and October 2014. We assigned a county to each tweet, then aggregated daily tweeting activity patterns for a typical weekday, and investigated to what extent do hourly activities correlate with employment or unemployment levels. We then modelled daily activity patterns as being the superposition of two universal patterns, thus aiming for a simple linear approximation of our dataset. By minimizing the squared error of our estimations, we obtained that the difference of the two patterns should be parallel to the first eigenvector of the covariance matrix of the dataset and that the mean of the data should fit on our line when selecting only 2 dimensions, and when using all 24 dimensions of our data as well. The set of eigenvalues of the covariance matrix in both cases confirmed the validity of our linear model, which captured most (0.99,0.52) of the variance in the dataset. Whereas in the 2-dimensional case the first eigenvector pointed to the direction, where 1pm activity was increased, and 0am activity decreased, in the 24-dimensional case it had positive elements during the daytime hours (6am-8pm), and was negative during the most of the night (9pm-5am).

By projecting county activity patterns onto these lines with the mean as the origin, we obtained a measure for each country that indicated the extent to which the tweeting pattern of a county resembles that of the first eigenvector. This measure has been shown to correlate significantly with county labor force shares and unemployment rates, though in the 2-dimension case, these correlations could not enhance the performance of the single hourly correlations.  Using all 24 dimensions, we obtained a better Pearson correlation of $0.46\pm 0.02$ and $-0.34\pm 0.02$ for employment and unemployment, respectively. The signs of the correlations indicate a relationship where counties exhibiting a higher tweeting activity during the daytime (6am-8pm) have higher employment and lower unemployment rates, and counties with increased night activity can be related to lower employment and higher unemployment rates. These correlations show, that even though Twitter population is biased towards younger age groups, and employment data was considered for all age groups, the underlying relationship between daily activity patterns and employment data can be captured with plausible outcomes.

Our results thus showed, that by analyzing a relatively sparse publicly available geolocated dataset, a very simple model can explain to a significant extent such an important socio-economic indicator as employment/unemployment. We believe that our model can be even further improved by incorporating detailed data for different age groups or other datasets from either traditional or digital sources such as mobile traffic data. It would be worth to investigate whether dynamic changes of activity patterns over time can follow employment trends. This kind of analysis would allow policy makers a better insight into the processes connected to employment phenomena, and could form the basis of future datasets, where problems could not only be identified based on officially registered unemployed people, but also on a basis of the digital footprints people leave on different platforms.

\clearpage

\section*{Figures}

\begin{figure}
	\begin{center}
		\includegraphics[width=0.3\tw]{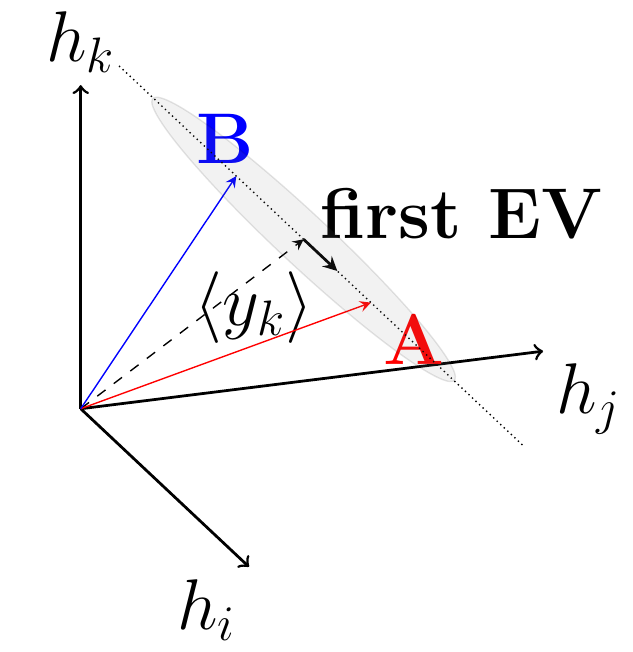}
	\end{center}
	\label{fig:vec}
	\caption{{\bf Schematic figure of linear model.} Linear representation of the datapoints by $\mathbf{A}$ and $\mathbf{B}$ corresponding to the two universal patterns later identified as the active and inactive patterns, the dashed line showing the mean value of activities. The bold vector is the direction of the first eigenvector of the covariance matrix $\mathbf{C}$. $h_i$, $h_j$ and $h_k$ represent three arbitrarily chosen axis corresponding to different hours $i$, $j$ and $k$ of the day.}

\end{figure}

\clearpage

\section*{Authors' contributions}

G.V. conceived the experiment, E.B. and Z.L. collected the data, E.B. and G.V. analyzed the results, E.B. wrote the manuscript. All authors reviewed the manuscript.

\section*{Availability of data and materials}

The dataset supporting the conclusions of this article is available in the following repository: \url{http://www.vo.elte.hu/papers/2016/unemployment/}.

\section*{Competing interests}

The authors declare that they have no competing interests.

\clearpage

% \bibliographystyle{spphys}
% \bibliography{biblio2}

\clearpage

\begin{flushleft}
{\Large
\textbf\newline{Technical details of the linear model calculations for the article\\
Prediction of employment and unemployment rates from Twitter daily rhythms in the US}
}
\newline
\\
Eszter Bokányi, bokanyi@complex.elte.hu\textsuperscript{*},\\
Zoltán Lábszki, labszkizoltan@gmail.com,\\
Gábor Vattay, vattay@complex.elte.hu\\
\bigskip
Department of Physics of Complex Systems\\
Pázmány Péter sétány 1/A, Eötvös Loránd University, Budapest H-1117, Hungary
\bigskip

* corresponding author
\end{flushleft}

\section*{Technical details for the Methods section}

We define a daily activity pattern with hourly resolution for each county that 
are enumerated by $k=1\dots M$. Thus, each county ($k$) is represented by a 24-dimensional vector ($\yk$), where the 
elements of $\yk$ are aggregated normalized hourly tweeting activities.

We assume that the tweeting pattern of a county can be represented by the linear combination of only two universal patterns ($\mathbf{A}$ and $\mathbf{B}$) that are mixed for each county $k$ with a proportion of $\alpha^{(k)}$ and $1-\alpha^{(k)}$, respectively. We have no further restriction on these $\alpha^{(k)}$ values, they can be any arbitrary real numbers. $\mathbf{A}$ and $\mathbf{B}$ are both 24-dimensional vectors normalized to 1, the 24 dimensions representing the 24 hours of the day.

Then the predicted activity $x_i^{(k)}$ of a county $k$ in hour $i$ would be
\begin{equation}
x_i^{(k)}=\alpha^{(k)}\cdot A_i+(1-\alpha^{(k)})\cdot B_i=\alpha^{(k)}(A_i-B_i)+B_i.
\label{eq:linear}
\end{equation}

Let us denote the weight of each county by $\wk$, which is proportional to its population $p^{(k)}$, such that $\wk=p^{(k)}/\sum_{k=1}^M p^{(k)}$. We then define the squared error of our model as
\[E=\sum_{i,k}\wk\left(y_i^{(k)}-\underbrace{\left(\alpha^{(k)}(A_i-B_i)+B_i\right)}_{x_i^{(k)}}\right)^2.\]

\clearpage

We would like to minimize this error with subject to the two conditions $\sum_i A_i=1, \sum_i B_i=1$, which leads to the following expression to minimize with Lagrange multipliers $\lambda_a$ and $\lambda_b$:

\begin{equation}
	E+\lambda_a\left(\sum_i A_i-1\right)+\lambda_b\left(\sum_i B_i-1\right)=\mathrm{min.}	
\end{equation}

The derivatives yield the following linear equation system:
\begin{align}
	\frac{\partial}{\partial A_j}&:&\sum_k 2\wk\left(y_j^{(k)}-\alpha^{(k)}(A_j-B_j)-B_j\right)\left(-\alpha^{(k)}\right)+\lambda_a&=0 \label{eq:sys1}\\
	\frac{\partial}{\partial B_j}&:&\sum_k 2\wk\left(y_j^{(k)}-\alpha^{(k)}(A_j-B_j)-B_j\right)\left(-(1-\alpha^{(k)})\right)+\lambda_b&=0 \label{eq:sys2}\\
	\frac{\partial}{\partial \alpha^{(m)}}&:&\sum_i 2\wm\left(y_i^{(m)}-\alpha^{(m)}(A_i-B_i)-B_i\right)\left(-(A_i-B_i)\right)&=0 \label{eq:sys3}
\end{align}

Summing Eq~\ref{eq:sys1} and Eq~\ref{eq:sys2} for $j$ yield 0 for the Lagrange multipliers $\lambda_a$ and $\lambda_b$. Thus, the problem reduces to minimizing $E$, which actually measures the sum of squared distances from the line parametrized by $\mathbf{A-B}$, $\mathbf{B}$ and $\alpha^{(k)}$ for a county $k$.

Since
\begin{equation}
	\sum_j [(\ref{eq:sys1})+(\ref{eq:sys2})]\cdot (A_j-B_j) = \sum_m (\ref{eq:sys3}),
\end{equation}
the equation system is not linearly independent. Thus, we cannot obtain all exact values for $A_j$, $B_j$ and $\alpha^{(k)}$, they will be dependent on each other.

Expressing $\alpha^{(k)}$ from our equation system yields:
\begin{equation}
\label{eq:alpha}
\alpha^{(m)}=\frac{\sum_i \left(y_i^{(m)}-B_i\right)\left(A_i-B_i\right)}{\sum_i\left(A_i-B_i\right)^2}=\frac{(\mathbf{y}^{(m)}-\mathbf{B})(\mathbf{A}-\mathbf{B})}{(\mathbf{A}-\mathbf{B})^2}.
\end{equation}

The line from which the summed distance of the datapoints is minimal is the line whose direction is parallel to the eigenvector ($\mathbf{m}$) corresponding to the largest eigenvalue of the covariance matrix $\mathbf{C}$, where
\begin{equation}
	C_{ij}=\left\langle y_iy_j\right\rangle-\left\langle y_i\right\rangle \left\langle y_j\right\rangle,
\end{equation}
if  $\left\langle\right\rangle$ denotes the weighted mean ($\sum_k \wk=1$, $\wk\ge 0\,\forall k=1\dots M$)
\begin{equation}
	\left\langle y_j\right\rangle=\sum_k \wk y_j^{(k)}.
\end{equation}

By substituting the expression for $\alpha^{(k)}$ into Eq~\ref{eq:sys1})+Eq~\ref{eq:sys2}, and averaging over $k$ we get that the point $\left\langle \mathbf{y}\right\rangle$ should fit onto our line.

\clearpage

Thus, we get a valid solution of our error minimization problem, if we choose
\begin{eqnarray}
	\sigma(\alpha)=\sqrt{\frac{\sum_{k=1}^M\left(\alpha^{(k)}\right)^2}{M}}, \nonumber\\
	\label{eq:a} \mathbf{A}=\left\langle \mathbf{y}\right\rangle + 2\cdot \mathbf{m}\cdot \sigma(\alpha),\\
	\label{eq:b} \mathbf{B}=\left\langle \mathbf{y}\right\rangle - 2\cdot \mathbf{m}\cdot \sigma(\alpha),
\end{eqnarray}
and calulate $\alpha^{(k)}$ values according to Eq~\ref{eq:alpha}.

\end{document}